# Anisotropic Strain Induced Soliton Movement Changes Stacking Order and Bandstructure of Graphene Multilayers

Fabian R. Geisenhof[1], Felix Winterer[1], Stefan Wakolbinger[1] Tobias D. Gokus[2], Yasin C. Durmaz,[3], Daniela Priesack[1], Jakob Lenz[1], Fritz Keilmann,[4], Kenji Watanabe[5], Takashi Taniguchi[5], Raúl Guerrero-Avilés[6], Marta Pelc[6], Andres Ayuela[6], R. Thomas Weitz[1,4,7,*]

## Abstract

The crystal structure of solid-state matter greatly affects its electronic properties. For example in multilayer graphene, precise knowledge of the lateral layer arrangement is crucial, since the most stable configurations, Bernal and rhombohedral stacking, exhibit very different electronic properties. Nevertheless, both stacking orders can coexist within one flake, separated by a strain soliton that can host topologically protected states. Clearly, accessing the transport properties of the two stackings and the soliton is of high interest. However, the stacking orders can transform into one another and therefore, the seemingly trivial question how reliable electrical contact can be made to either stacking order can a priori not be answered easily. Here, we show that manufacturing metal contacts to multilayer graphene can move solitons by several µm, unidirectionally enlarging Bernal domains due to arising mechanical strain. Furthermore, we also find that during dry transfer of multilayer graphene onto hexagonal Boron Nitride, such a transformation can happen. Using density functional theory modeling, we corroborate that anisotropic deformations of the multilayer graphene lattice decrease the rhombohedral stacking stability. Finally, we have devised systematics to avoid soliton movement, and how to reliably realize contacts to both stacking configurations.

[1] Physics of Nanosystems, Department of Physics, Ludwig-Maximilians-Universität München, Amalienstrasse 54, 80799 Munich, Germany
[2] neaspec GmbH, Bunsenstrasse 5, Martinsried, 82152 Munich, Germany
[3] Department of Physics, Ludwig-Maximilians-Universität München, Schellingstr. 4, 80799 Munich, Germany
[4] Center for Nanoscience (CeNS), Schellingstr. 4, 80799 Munich, Germany
[5] National Institute for Materials Science, Tsukuba, Japan
[6] Donostia International Physics Center (DIPC), Paseo Manuel Lardizabal 4, 20018 Donostia-San Sebastián, Spain
[7] Nanosystems Initiative Munich (NIM), and Munich Center for Quantum Science and Technology (MCQST), Schellingstrasse 4, 80799 Munich, Germany
*Correspondence and requests for materials should be addressed to R.T.W., thomas.weitz@lmu.de

Recent interest in graphene multilayers stems from their diverse (opto-)electronic properties that depend on layer thickness[1], stacking order[2–6] and twist angle of subsequent layers against one another[7,8]. For example, Bernal-stacked bilayer graphene shows an electrically tunable bandgap[9], unconventional quantum-[10] and fractional-quantum Hall effects,[11–13] as well as a renormalization of the density of states near charge neutrality in the absence of a magnetic field[14–17]. Furthermore, it has been recently shown that slightly twisting the two layers dramatically changes the band structure, allowing the observation of unconventional superconductivity[8] and Mott insulating behavior[18]. An addition to these diverse properties of graphene bilayers are the recently identified topologically protected states at boundaries between AB and BA stacked graphene bilayers[19–22]. In thicker graphene flakes, the physics can be expected to be even richer, since naturally two stable forms of stacking, Bernal (or ABA, Fig. 1a) and rhombohedral stacking (or ABC, Fig. 1b) exist, both with distinct bandstructures and electronic properties[2–4]. For example, in thick rhombohedral graphene stacks flat electronic bands are present[23,24], which might support superconductivity[25–28].

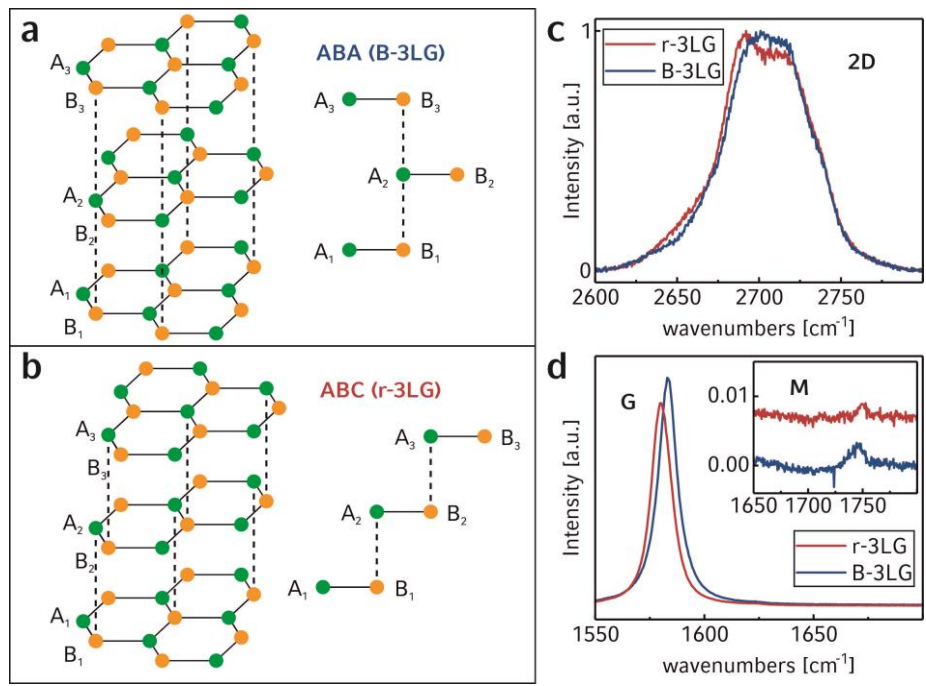

*Figure 1 | Crystal structure and Raman modes for trilayer graphene with Bernal (ABA) and rhombohedral (ABC) stacking order. a,b, Crystal structure (left) and cross section (right) of trilayer graphene with Bernal (a) and rhombohedral (b) stacking. c,d, Raman spectra in trilayer graphene showing the 2D (c), G (d) and M mode (d inset) for Bernal (B-3LG, blue) and rhombohedral stacking (r-3LG, red). The 2D peaks are normalized. The M band is normalized to the G mode intensity and an offset is used for better visibility. An AFM image of the flake on which the spectra were acquired is shown in Fig. 3.*

To explore the unique electronic properties of both stacking orders, precise knowledge of the local stacking type as well as stable electric contacting of both phases is necessary. This is even more important, since both stacking orders can occur within the same flake. Moreover, as recently shown, both types of stacking can be transformed into one another by applying an electrical field[29], strain[30], high temperatures[31], doping[32] or a mechanical force[33] – some of which are also present during the patterning of electrical contacts or the transfer of multilayers onto hexagonal Boron Nitride (h-BN). With the goal to assess the stability of the stacking orders under processing and to devise reliable contacting schemes, we have investigated a series of multilayer graphene flakes.

We have organized this manuscript into three sections. First, we investigate the stability of the stacking orders under metal contact patterning. Second, we focus on the stability of both phases on h-BN. In the final section we discuss how to avoid soliton movement during processing.

1. **Stability of rhombohedral and Bernal stacking under metal contact processing**

**Characterization of multilayer graphene before and after processing.** Before electrically contacting our graphene multilayers, we identify the local stacking order using Raman spectroscopy of the 2D, G and M mode (Figs. 1c and d) as previously shown[34–37]. Since within a single multilayer graphene flake Bernal and rhombohedral stacking order can be present, we spatially resolve the stacking domains using scanning Raman spectroscopy[35]. Fig. 2 shows in addition to an AFM (Fig. 2a) and optical image (Fig. 2b) a map of the full width at half maximum (FWHM) of the 2D peak of a tetralayer graphene flake after exfoliation (Fig. 2c), that reveals two different stacking domains. Knowledge of the local stacking order allows us to selectively pattern contacts on the Bernal and rhombohedral domains *via* standard electron beam lithography, metal evaporation and lift-off techniques. An optical image of the flake with fully processed contacts is shown in Fig. 2e and an AFM image in Fig. 2d. In the latter, new wrinkles between the contacts can be observed. The appearance of such folds can be caused by compressive and/or shear strain[38–40], induced by thermal expansion effects[40,41].

At first sight, the appearance of strain and the occurrence of folds upon deposition of metal contacts does not seem worth a separate discussion, especially since numerous previous works have reported metal contacts to both rhombohedral and Bernal stacked multilayer graphene[2,3,5,42,43]. Nevertheless, we have investigated the contacted flake again with scanning Raman spectroscopy. Surprisingly, we find that the rhombohedral domain has almost completely vanished (see Fig. 2f), even though it was covering an area of about 40 µm² prior to contact deposition. From the Raman 2D signal we can clearly identify that it has transformed to Bernal stacking[36] (see Fig. 2i), corroborated by spectra of the G (Fig. 2j) and the M mode[36] (Fig. 2j, inset). Finally, the D peak is negligible indicating that the transition has not introduced defects. Since the lateral resolution of our Raman microscopy setup is about 1 µm, we cannot rule out that the rhombohedral part of the flake might have been broken up into several nanodomains of Bernal and rhombohedral stacking[44]. We have therefore investigated the local nature of the transformed flake by infrared scattering-type scanning near-field optical microscopy (s-SNOM)[45] at 20 – 30 nm spatial resolution. This technique is highly suitable in our case, as it allows (i) to distinguish Bernal from rhombohedral stacking owing to their different infrared responses[19,33,46,47], and

(ii) to resolve boundaries between stacking domains in multilayer graphene (*e.g.* boundaries between ABA – BAB or ABA – ABC) due to reflections of surface plasmons[19,48,49]. An s-SNOM image of the processed flake (Fig. 2h) confirms the spatial arrangement of the domains, as revealed by the Raman map (Fig. 2f). With its high resolution, the s-SNOM image corroborates that the transition from rhombohedral to Bernal stacking proceeds homogenously upon the contacting process. In other words, the transition does not nucleate at multiple points within the flake, but seems to be induced *via* movement of the strain soliton at the Bernal/rhombohedral stacking boundary. For a better visualization of the transition, a scheme including the rhombohedral, Bernal and transformed region is depicted in Fig. 2g.

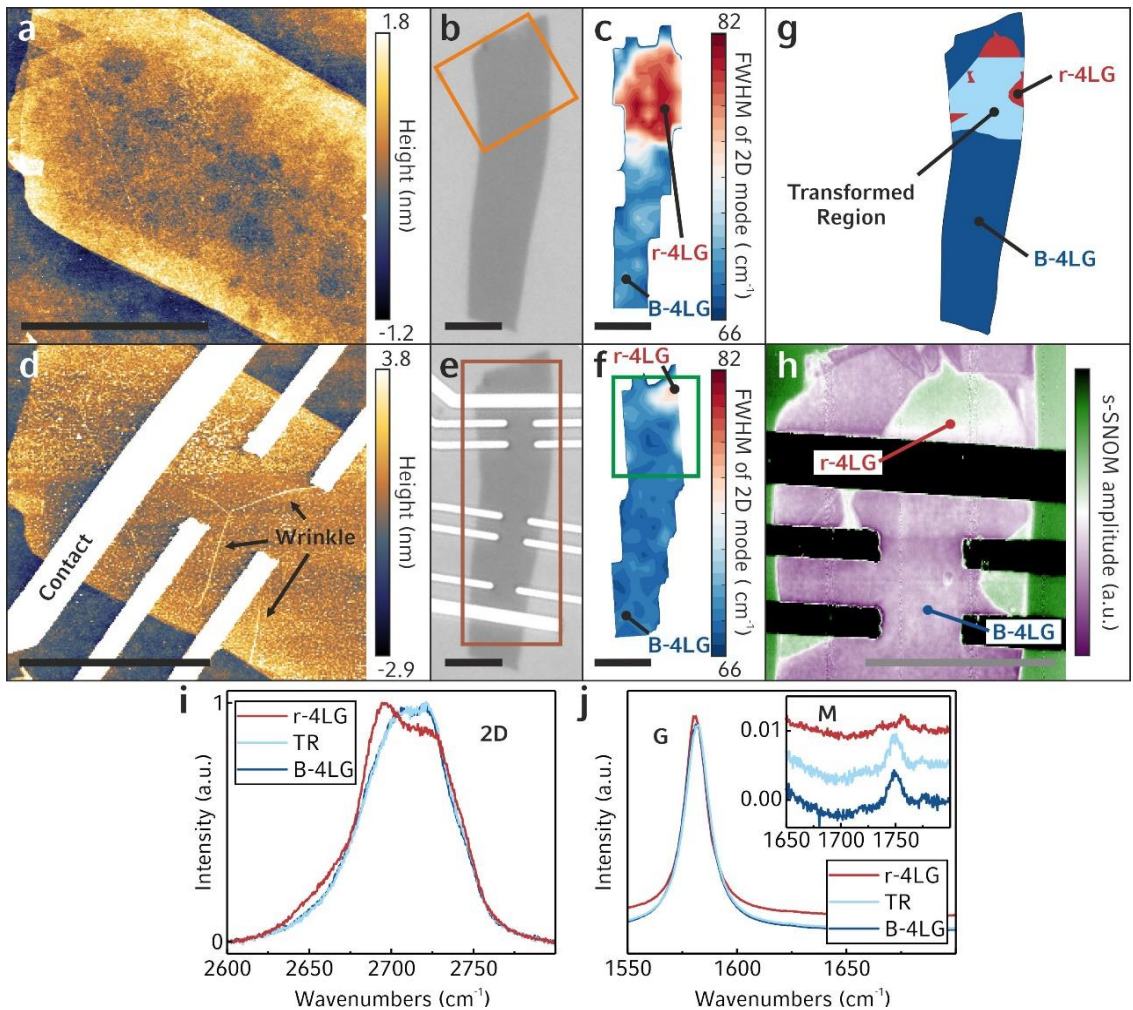

**Figure 2 | Stacking transformation in a graphene tetralayer observed by Raman spectroscopy and infrared s-SNOM nano-imaging.** *a,d, AFM image of the top part of a tetralayer graphene flake before (a) and after (d) fabricating contacts. b,e, Optical image of the pristine (b) and processed (e) tetralayer, the orange and brown rectangles indicate the regions shown in (a),(d), and (c),(f), respectively. c,f, Map of the FWHM of the 2D mode in the pristine (c) and processed (f) flake. The rhombohedral and Bernal stacking domains are indicated. The green rectangle denotes the region shown in (h). g, Schematic representation of the arrangement of the stacking domains in the processed flake. h, s-SNOM amplitude image in the top part of the processed flake. Two domains of different stacking order are visible, with sharp boundary. i,j, Raman spectra recorded in the processed flake in the rhombohedral (r-4LG, red), Bernal (B-4LG, blue) and transformed region (TR, light blue) showing the 2D (i), G (j) and M mode (j, inset). The 2D peaks are normalized. The M band is normalized to the G mode intensity and an offset is used for better visibility. Scale bar in all images: 5 μm.*

This transformation from rhombohedral to Bernal stacking has been observed in about 50 % of the contacted samples, which are flakes with 3 to 7 graphene layers. The extent of the transition varies, ranging from an almost complete vanishing of the rhombohedral domain as in Fig. 2, to only a small movement of the domain wall by a few hundred nm. In flakes in which no transition occurs, the soliton might be pinned, as observed before by STM measurements[29]. Further below, we discuss additional reasons why in some cases the soliton does not move.

**Identifying the cause of the stacking transformation.** We have made an attempt to clarify the detailed mechanism that causes the transition, with the aim to devise a way to avoid the transition or enhance it selectively. The most remarkable observation next to the stacking transition in the flake shown in Fig. 2 is the occurrence of wrinkles in the transformed region. This might imply that the transformation is directly correlated to or caused by the appearance of wrinkles. However, such wrinkles do not necessarily appear in the parts of the flake in which a transition takes place. For example, in Fig. 3, we show details of a graphene trilayer that comprises a Bernal/rhombohedral stacking boundary. After fabricating metal contacts, as in the case of the tetralayer, the high-resolution s-SNOM image (Fig. 3g) shows that again the domain boundary has shifted, slightly increasing the Bernal-stacked region, whereas the wrinkles occur primarily in the untransformed rhombohedral part. This implies that both the stacking transformation and the topography changes appear simultaneously during processing and possibly originate from the same cause, however, the folds themselves do not trigger the transition.

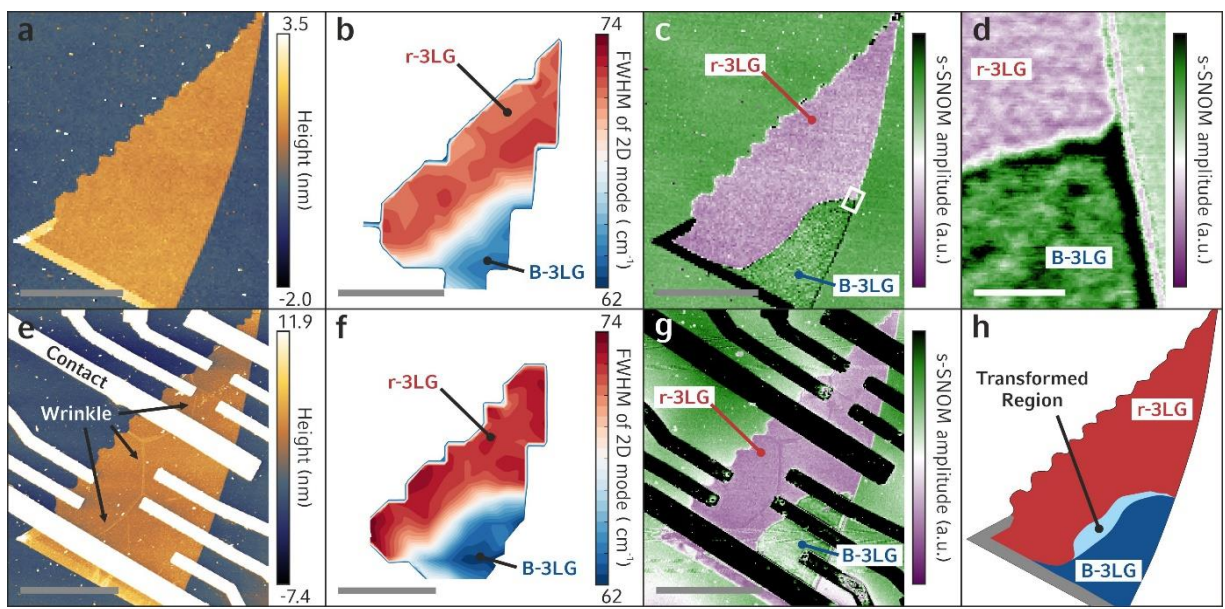

*Figure 3 | Soliton movement in trilayer graphene. a,e, AFM image of a pristine (a) and processed (e) trilayer graphene flake. b,f, Spatial map of the width of the 2D mode recorded in the pristine (b) and processed (f) flake. The rhombohedral and Bernal stacking domains are indicated. Individual Raman spectra of the flake are shown in Fig. 1c and d. c,g, s-SNOM image showing the optical amplitude measured in the pristine (c) and processed trilayer (g). The white rectangle in (c) denotes the region shown in (d). The domains of different stacking order are marked. d, Zoomed-in image of the sharp stacking boundary. h, Scheme of the stacking domains in the processed flake. The rhombohedral, Bernal and transformed regions are indicated in red, blue and light blue, respectively. Scale bar: 5 μm in (a-c) and (e-g), 300 nm in (d).*

To identify the cause of the transition, we apparently need to test if during our process steps effects occur that can lead to a stacking transformation, with special attention to mechanisms that are also known to induce folds. A few methods including applying an electrical field[29], strain[30], high temperatures[31], doping[32], an electron beam[22] and recently also a mechanical force[33] have been reported to cause a movement of solitons or a transformation between Bernal and rhombohedral graphene or vice versa. Since we do not apply an electric field across the flake, we can exclude this directly as possible trigger for a transformation. We next turn to doping as possible cause. In principle, metals, when in close contact with graphene, can lead to doping[50]. However, our contacts are deposited locally at the edges of our multilayers, and screening lengths are well below 100 nm[51], whereas the transition occurs non-locally across several µm. Furthermore, we have deposited a few nm of titanium onto several multilayer flakes with both forms of stacking and have not observed a transition. Finally, the contacts seem to rather hinder than foster the movement of the domain wall (see Fig. 3 and Supplementary Fig. S5). Additionally, we have corroborated that an electron beam does not cause the transition (for details see Supplementary Fig. S1). Consequently, strain and high temperature are left as possible explanations. Since both can possibly occur during processing, we subsequently investigated the processing steps of cleaning, heating, spin-coating PMMA and softbake and identified that they do not cause a transformation (see Supplementary Fig. S1). Thus, we conclude that the metal evaporation is the decisive step, the effect of which we have investigated in the following.

During the electron beam evaporation of metals, the substrates are held at about 10 – 20 °C by water-cooling of the sample holder. Still, the graphene flakes can heat up locally caused by the thermal load of the condensing metals. Consequently, due to the large difference in the thermal expansion coefficients of graphene (about -8.0 x $10^{-6}$ 1/K, ref. 52), PMMA (about 100 x $10^{-6}$ 1/K, ref. 53) and the SiO$_2$ substrate (about 0.6 x $10^{-6}$ 1/K, ref. 54), mechanical strain can occur[40]. As previously reported, the stress transfer between graphene and PMMA is very good for small strain values[55,56], however, at strain of 0.6 % or higher, slippage between the two materials can occur[55,56]. In the case of PMMA homogeneously covering a multilayer graphene flake (see Fig. 4a), upon heating and subsequent cooling the resist, the multilayer experiences homogeneous tension and/or compression, depending on whether slippage has occurred or not. Given the typical temperatures reached in our experiment (180 °C in the case of the PMMA softbake), the reached homogeneous expansion of PMMA and the resulting tension/compression of graphene is below 1.6 %. We have never observed a change in stacking order under these circumstances, which fits our DFT calculations (see Supplementary Fig. S2) that upon isotropic strain the relative stability of both stacking orders does not change. The situation is different in the case that e-beam lithography has been performed on the PMMA resist, as shown in Fig. 4b. Even though the detailed geometry of the sample with contacts is complicated, it seems plausible that during evaporation the flake and the resist heat up, with the latter expanding thermally and therefore stretching the flake (where parts of the flake might have slipped with respect to PMMA). When the metal forms a closed layer, it locally pins the flake in the heated condition. The subsequent cool-down leads to a contraction preferably in the part of the flake not pinned by the contacts and therefore, to anisotropic strain in the flake.

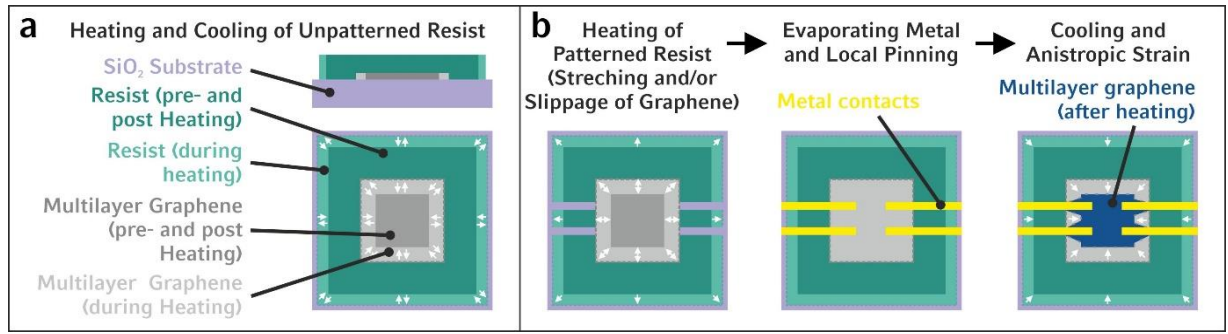

*Figure 4 | Suggested mechanism of transformation. a, Schematic illustration of heating and cooling the sample covered with unpatterned resist leading to homogeneous deformation. The arrows indicate the heating and subsequent cooling of the sample. b, Proposed mechanism of how the combination of heating, local pinning during processing and subsequent cooling leads to anisotropic strain causing the preference of Bernal stacking. Left: the patterned resist heats up during evaporation and expands thermally, streching the flake. Middle: The metal is forming a closed layer, locally pinning the flake in the hot state. Right: The resist cools down, contracting the flake in the non-pinned part of the flake thus inducing anisotropic strain in the flake.*

To understand the combined effect of heating and mechanical strain during metal evaporation and to test if we can amplify the scale of the transformation in experiment, we deliberately increased the substrate temperature during metal deposition to 200 °C. AFM and s-SNOM images before and after metal evaporation at 200 °C and lift-off are shown for a tetralayer graphene flake in Figs. 5a – d (corresponding Raman data in Supplementary Fig. S3). Since 200°C is well above the glass transition of PMMA[57], one can assume that the corners of the patterned PMMA get softened and we consequently find that the metal contacts are torn off during the lift-off, resulting in an inhomogeneous surface of the processed flake (see Fig. 5c). Nevertheless, our measurements after processing show new wrinkles in the topography (see Fig. 5c) and a full transformation to Bernal stacking (see Fig. 5d). Apparently, the increase of the substrate temperature during metal evaporation has amplified the scale of the transformation in case that additionally the resist has been patterned.

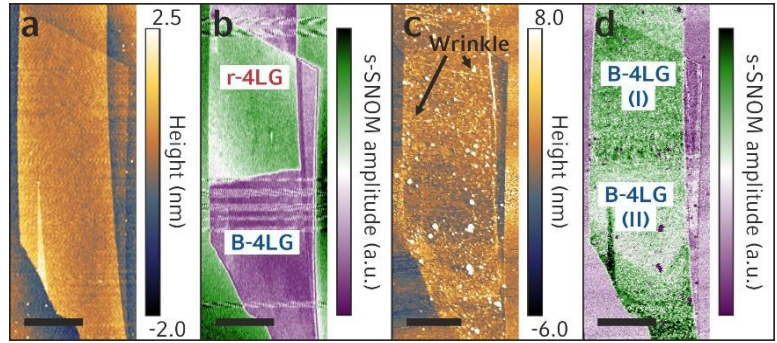

*Figure 5 | Forcing soliton movement. a,c, AFM image of a pristine (a) and processed (c) tetralayer graphene flake. b,d, s-SNOM images showing the optical amplitude measured in the pristine (b) and processed tetralayer (d). In b, domains with rhombohedral and Bernal stacking are marked. In d, two sections (part I and II) are indicated and explained in the SI. The ratio of the signal of the flake and the substrate changed compared to the ratio in the pristine flake, probably due to remaining residues of the resist. Scale bar in all images: 2 μm.*

We also note that the anisotropic strain variations which are per se present in graphene flakes on $SiO_2$[58] are apparently not strong enough to induce the transition, since we never have observed that heating the flake (even in the case that PMMA has been deposited on the flake) has induced a transition. Therefore, we anticipate that the combination of thermal heating and local clamping of parts of the flake with metal contacts causes anisotropic compressive strain and shear forces, which then lead to the observed folds[38–40] and, more importantly, provides a driving force for the soliton movement causing the transition. This is corroborated with the results of our DFT calculations that under anisotropic strain the energy of rhombohedral stacking rises faster than the energy of Bernal stacking (see Supplementary Fig. S2). Consequently, the relative stability of Bernal stacking increases under anisotropic lattice deformations, making it more favorable. Our explanation is in line with the recent observation by de Sanctis et al.[59] that fabricating contacts can induce a strain pattern and also agrees with our observation that the extent of the transition can vary or not occur at all, since the occurring temperatures, the design of the contacts and the resulting strain during the fabrication of metal contacts can be different for each flake. Since we do not observe a significant shift of the 2D Raman mode in the finished flakes[55,56], we conclude that after the resist has been removed, the graphene is unstrained – most probably due to the appearance of folds.

## 2. Stability of rhombohedral and Bernal stacking under dry transfer onto h-BN

In the case that a high sample quality is required, multilayer graphene is frequently either placed on top of or embedded within h-BN flakes[60,61]. For rhombohedral graphene multilayers this is however not straight forward and, it has been reported previously[6] that during the encapsulation process rhombohedral stacking can transfer to Bernal stacking. It would be therefore interesting to identify during which step of the encapsulation process the anisotropic strain is present. We therefore have used the PDMS all-dry viscoelastic stamping method[62] to bring a r-3LG flake onto h-BN. To this end, we have exfoliated graphene directly onto PDMS stamps and identified suitable graphene multilayers via their optical contrast. Subsequently, we have transferred the flake onto h-BN (see Fig. 6a for an AFM image), and performed a detailed map (Fig. 6b) of the 2D Raman peak (Fig. 6c). As can be seen, parts of the flake show rhombohedral stacking, while other parts display Bernal stacking. Furthermore, one can recognize (see Fig. 6d) that there is a layer of contaminants between the multilayer graphene flake and the h-BN layer – since the transfer process is performed in ambient environment. To benefit from the flat and clean interface between the graphene multilayer and h-BN, and to avoid high temperature annealing, we have then used our recently developed method of bringing the multilayer graphene flake into close contact with h-BN with an AFM tip to remove contaminants between the layers[63]. This local point of close contact and mild heat treatment up to 40°C is enough to remove the entire contaminants at the interface. After the cleaning (see Fig. 6e), the flake has transferred completely to Bernal stacking (Fig. 6f, g and h). It is reasonable to assume that also here anisotropic strain is driving the transformation, since in the cause of the cleaning process certain regions of the TLG will touch the h-BN first in close van-der-Waals contact while other parts will still be separated by the layer of contaminants. In this state, most probably anisotropic strain is present causing the transformation. To make samples of an even higher quality, one typically fully encapsulates the graphene flakes in h-BN. It is reasonable to assume, that also here during the pick-up process of the flakes in the moment when part of the graphene flake is already in close contact with the h-BN flake, and parts of it still rest on the substrate anisotropic stress is present and induces the transition.

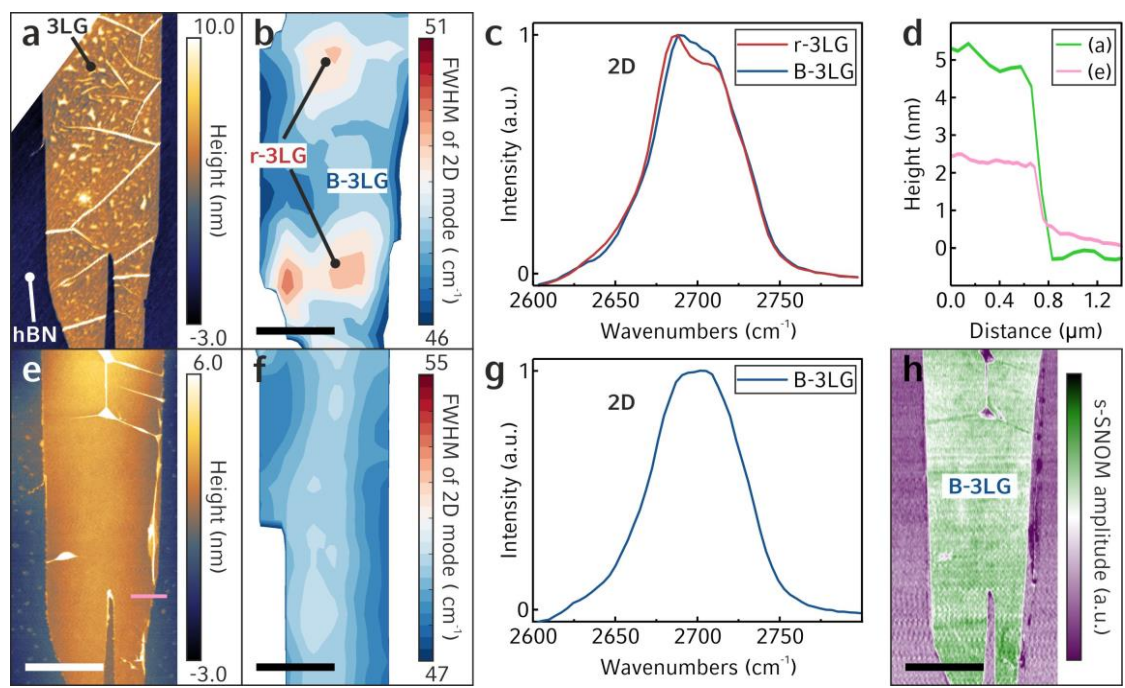

*Figure 6 | Soliton movement upon transfer of multilayer graphene onto h-BN. a,e, AFM image and b,f, scanning Raman maps of the 2D peak of a graphene multilayer right after PDMS stamping (a,b) and after the flake has been brought into intimate contact with h-BN (e,f). c,g Corresponding individual Raman spectra of the 2D peak before (c) and after (g) bringing the flake into close contact. d, Linecuts of the AFM images shown in (a,e). h, s-SNOM image showing the optical amplitude of the flake in close contact to hBN showing that it has transformed homogeneously to Bernal stacking. Scale bar in all images: 3 μm.*

## 3. Stability of finished samples and methods to avoid soliton movement

After having established the different scenarios in which a transformation from rhombohedral to Bernal stacking in multilayer graphene flakes can happen, we next asses if the transformed regions are stable under typical measurement conditions. To this end, we performed *ex situ* measurements with the transformed samples. We cooled the samples in liquid nitrogen or helium and heated the contacted samples to 400 °C and performed AFM, Raman and s-SNOM measurements afterwards (see Supplementary Fig. S4). The folds moved, changed height or disappeared but the arrangement of the stacking domains remained unchanged.

**How to avoid soliton movement during contacting.** While we have now established heating the sample during metal deposition or establishing close van-der-Waals contact with h-BN as a method to induce soliton movement, to access the physics of rhombohedral domains or of states at the rhombohedral-to-Bernal-stacking boundary, it would be beneficial to devise a way to avoid soliton movement. To this end, the applied strain during processing needs to be kept as small as possible. This can be achieved by, firstly, assuring that the sample is cooled well during the evaporation, since a higher temperature amplifies the scale of the transition, and secondly, by choosing the right pattern of contacts. The latter can help to prevent a transformation, since the contacts seem to hinder the movement of the soliton. During the transition, the soliton at the Bernal/rhombohedral stacking boundary moves towards the rhombohedral part. If the contacts lie across the domain wall, the soliton shifts only slightly in between the contacts (see Fig. 3), thus, these contacts prevent a free movement of the soliton. In case a contact fully separates the rhombohedral domain from the Bernal part, the transformation stops (see Supplementary Fig. S5). Finally, we have found that a dense contact pattern around the edges of the flake effectively suppresses soliton movement, preventing any stacking transition (see Supplementary Fig. S6). The high density of metal contacts clamps the flake on all sides effectively and the anisotropy of the strain is consequently reduced, causing no transition. Finally, we have shown that e-beam and resist deposition do not induce a movement of the strain soliton and furthermore, the transition is always initiated from Bernal stacked regions. This implies, that in the case multilayer graphene shall be encapsulated in h-BN, a safe way to avoid transition to Bernal stacking should be to remove all non-rhombohedral parts via etching prior to transfer. It seems that the transfer is the only critical step in the encapsulation and contacting process, since it had been shown previously, that the frequently used 1D side contacts do not cause anisotropic strain in graphene[59], i.e. most probably do not cause a transition.

In summary, we have observed that metal contact patterning or when multilayer graphene is brought into intimate contact with h-BN can induce – most probably due to anisotropic strain - a movement of stacking solitons at the rhombohedral/Bernal boundary in multilayer graphene. Even though it is known that graphene layers can easily move against one another[64], the here reported transition is surprising since numerous literature reports have contacted rhombohedrally stacked multilayer graphene without reporting such a transition[2,3,5,42,43]. The observation that fabricating metal contacts can lead to a lateral movement of van-der-Waals multilayers will also potentially be interesting for van-der-Waals heterostructures, where atomic lateral precision or twist angles are required[8].

## Methods

**Sample preparation.** We mechanically exfoliated multilayer graphene flakes from an HOPG block onto a $SiO_2$(300 nm)/Si substrate. The number of layers was determined by optical microscopy and Raman spectroscopy. The AFM images were recorded using an AFM (Dimension 3100, Veeco) in tapping mode. For the heterostructures, the graphene flakes where directly exfoliated onto PDMS stamps. The h-BN was synthesized[65] and exfoliated directly onto $SiO_2$ substrates. The heterostructures were created following the all-dry viscoelastic stamping method[62].

**Electron beam lithography.** The samples were cleaned using acetone and isopropanol. Then, the resist (PMMA 950K with 4.5 % anisole, Allresist) was spin coated onto the substrates. A softbake was performed at 180 °C for 5 min. The resist was patterned using an electron beam (e-Line system, Raith). Afterwards, the resist was developed using a 1:3 mixture of MIBK and isopropanol. Finally, the metals were deposited using e-beam evaporation under high vacuum conditions (pressure about $10^{-7}$ mbar) while the substrates were cooled (10 – 20 °C). The experiments in which the samples are heated to 200°C during evaporation were performed in a thermal evaporator. For all samples, a thin titanium layer of about 1 nm was applied as adhesion layer. Subsequently, gold was deposited forming the 30 – 80 nm thick contacts. The evaporation rates were about 0.1 Å/s and 1.0 Å/s, respectively.

**Raman measurements.** The spectra were recorded using a Raman system (T64000, Horiba) with a laser excitation wavelength of 514 nm. The size of the laser spot on the sample was about 1 µm and the spectral resolution was 0.7 cm$^{-1}$ (using a 1800 grooves/mm grating). The power of the laser spot was kept well below 1 mW to avoid local heating. The silicon peak at 521 cm$^{-1}$ was used as reference for wavenumber calibration. In order to get a spatial resolution of the stacking domains, the method described by Lui *et al.*[35] was used. The 2D mode is recorded every 1 µm. Then, a single Lorentzian peak is fitted to each spectrum and the FWHM is plotted. The spectra are background corrected to suppress the signal from nearby gold contacts. This procedure is further explained in the SI (see Supplementary Fig. S7).

**Infrared nano-imaging.** The infrared nano-imaging was performed using a commercial scattering-type scanning near-field microscope (s-SNOM, neaspec GmbH). Operating in intermittent contact AFM mode, topography and infrared nano-images of the graphene samples are obtained simultaneously.

For infrared nano-images, an infrared $CO_2$ laser beam with a wavelength of about 10.5 µm is focused onto a metal-coated AFM tip (Pt/Ir, Arrrow NCPT-50, Nanoworld). The tip oscillation frequency and amplitude were about 250 – 270 kHz and 50 – 80 nm, respectively. Acting as a nano-antenna, the AFM tip converts the incident infrared beam into a highly localized and enhanced electromagnetic field that is confined to its apex. This nanofocus creates a near-field interaction in the graphene underneath, whose magnitude depends on the local dielectric properties/optical conductivity of the graphene and thus is sensitive to layer number, stacking order and twist angle[19,33,46,47]. The near-field information is extracted from radiation back-scattered to a HgCdTe detector. We have also tested all-electronic Terahertz nanoscopy[66] at 0.6 THz to map the local stacking order, but did not observe any difference between rhombohedral and Bernal stacking (see Supplementary Fig. S8).

**Conflict of Interests:** T.D. Gokus, Y.C. Durmaz are an employee, and F. Keilmann is a cofounder of neaspec, producer of the s-SNOM microscope used in this study.

**Supporting Information Available:** Investigation of the first four processing steps; DFT calculations of the stability of stacking orders in trilayer graphene under lattice deformations; detailed Raman spectroscopy measurements for the tetralayer of Fig. 5; *ex situ* measurements showing the stability of the transformed regions; preventing a transition using a special arrangement of the contacts or a very dense contact pattern; details of the background correction used for the Raman spectra; comparing THz and infrared s-SNOM nano-imaging.

**Acknowledgement.** F.R.G., F.W., D.P., J.L., and R.T.W. acknowledge funding from the excellence initiative Nanosystems Initiative Munich (NIM), the Center for Nanoscience (CeNS) and the Solar Technologies go Hybrid (SolTech) initiative. We additionally acknowledge funding by the Deutsche Forschungsgemeinschaft (DFG, German Research Foundation) under Germany's Excellence Strategy – EXC-2111– 390814868 (MCQST) and EXC 2089 /1 -390776260." (e-conversion). We also thank Leonid S. Levitov, Nicola Mazzari and Nicolas Mounet for discussions and Jochen Feldmann for using his scanning Raman setup. R.G.-A., M.P. and A.A. thank the Project FIS2016-76617-P of the Spanish Ministry of Economy and Competitiveness MINECO, the Basque Government under the ELKARTEK project (SUPER), and the University of the Basque Country (Grant No. IT-756-13) for partial funding of this work. K.W. and T.T. acknowledge support from the Elemental Strategy Initiative conducted by the MEXT, Japan, A3 Foresight by JSPS and the CREST (JPMJCR15F3), JST.